\documentclass[%
reprint, 
tightenlines, 
superscriptaddress, 
showpacs, 
nobibnotes, 
amsmath, amssymb, 
aps, 
pra,
floatfix, 
dvipdfmx
]{revtex4-1}

\usepackage{latexsym}  
\usepackage{graphicx}
\usepackage{dcolumn}
\usepackage{bm}
\usepackage[usenames]{color} 
\usepackage{ulem} 
\usepackage{multirow} 
\usepackage{booktabs}

\allowdisplaybreaks

\begin{document}

\preprint{APS/123-QED}

\title{Nonholonomy of order parameters and su($3$) vortices in spin-$1$ Bose-Einstein condensates}

\author{Emi Yukawa} 
\email{emi.yukawa@riken.jp} 
\affiliation{RIKEN Center for Emergent Matter Science, 2-1, Hirosawa, Wako-shi, Saitama, 351-0198, Japan} 
\author{Masahito Ueda} 
\affiliation{Department of Physics, University of Tokyo, 7-3-1, Hongo, Bunkyo-ku, Tokyo, 113-8654, Japan} 
\affiliation{RIKEN Center for Emergent Matter Science, 2-1, Hirosawa, Wako-shi, Saitama, 351-0198, Japan} 

\date{\today}

\begin{abstract}
A generalized Mermin-Ho relation for a spin-$1$ BEC is derived, which is applicable to vortices regardless of the symmetry and spin polarization of the order parameter.  
The obtained relation implies an su(3) mass-current circulation and two classes of vortices corresponding to two different su($2$) subalgebras.  
\end{abstract}

\maketitle 

\section{\label{sec:1}Introduction}
In a superfluid with internal degrees of freedom such as superfluid helium-$3$ and spinor Bose-Einstein condensates (BECs), supercurrents are accompanied by spatio-temporal variations of spin and nematicity as well as the U($1$) phase~\cite{Salomaa1,Salomaa2}. 
Here, the rotation of the superfluid velocity, in general, does not vanish but depends on the spin-nematic texture, reflecting the nonholonomic nature of the texture. 
For the case of the $A$ phase of superfluid $^3$He ($^3$He-$A$), the nonholonomy leads to the celebrated Mermin-Ho relation~\cite{Mermin2}. 
The Memin-Ho relation is expressed in terms of three generators of the underlying so(3) symmetry of the order parameter, and implies that the circulation of the superfluid velocity is quantized when the loop encloses certain types of vortices~\cite{Salomaa1,Mermin1} such as the Mermin-Ho (MH) vortex~\cite{Mermin2} and the Chechetkin-Anderson-Toulouse (CAT) vortex~\cite{Chechetkin,Anderson}. 

In this paper we investigate the corresponding relation for spin-$1$ BECs. 
A new feature for this system is that not only the direction but also the magnitude of the spin vector can change over space and time and the spin nematicity arises as a consequence. 
We generalize the Mermin-Ho relation so as to be applicable to spin-$1$ BECs. 
The obtained relation involves eight generators and the corresponding structure constants of the su($3$) algebra, which implies the existence of vortices that belong to the other su($2$) subalgebra of the su($3$) algebra rather than the ordinary one corresponding to the above-mentioned so($3$) symmetry. 

\begin{figure}[t]
	\begin{center} 
		\includegraphics[bb = 0 0 1214 473, clip, scale = 0.175]{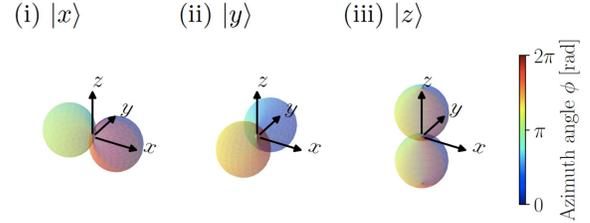}
	\end{center} 
	\caption{(Color Online) Three basis functions corresponding to $|x\rangle$, $|y\rangle$, and $|z\rangle$ in the Cartesian representation plotted in the spherical coordinate $(r, \theta , \phi)$, where $r$, $\theta$, and $\phi$ represent the radial coordinate, the polar angle, and the azimuth angle, respectively. 
	Here, $\phi \in [0, 2\pi )$, which is indicated by a rainbow spectral gradient, and is defined as the angle between the $x$ axis and the vector projected onto the $xy$ plane. 
	The basis functions are represented in terms of the rank-$1$ spherical harmonics as $u_x(\theta ,\phi ) := \langle \theta,\phi | x \rangle = ( - Y_1^1 (\theta ,\phi ) + Y_1^{-1} (\theta ,\phi ) ) / \sqrt{2}$, $u_y(\theta ,\phi ) := \langle \theta,\phi | y \rangle = i ( Y_1^1 (\theta ,\phi ) + Y_1^{-1} (\theta ,\phi ) ) / \sqrt{2}$, and $u_z(\theta ,\phi ) := \langle \theta,\phi | z \rangle = Y_1^0 (\theta ,\phi )$. 
	The radial coordinate is given by $r(\theta ,\phi ) = |u_{\mu} (\theta ,\phi )|$. 
	The color gauge shown on the right indicates the value of $\phi$.} 
	\label{fig:1}
\end{figure} 
In the mean-field description of a spin-$1$ BEC, all bosons condense into a single spin state as well as a single spatio-temporal mode. 
To describe the spin degrees of freedom, we adopt the Cartesian basis $\{ |\mu \rangle | \mu = x, y, z \}$ satisfying $F_{\mu} |\mu \rangle = 0$~\cite{Ohmi}, where $F_{\mu}$ is the $\mu$ component of the spin-vector matrix. 
In the Cartesian representation, the spin matrices are given by $(F_{\mu})_{\nu \lambda} = - i {\epsilon}_{\mu \nu \lambda}$, where ${\epsilon}_{\mu \nu \lambda}$ is the completely antisymmetric tensor. 
The basis state $|\mu \rangle$ can be expressed in terms of the eigenstates $\{ |m\rangle | m = 1, 0, -1 \}$ of $F_z$, namely the basis states of the irreducible representation, as 
\begin{align}
	& |x \rangle = \frac{1}{\sqrt{2}} \left ( - |1 \rangle + |-1 \rangle \right ), \label{eq:|x>} \\ 
	& |y \rangle = \frac{i}{\sqrt{2}} \left ( |1 \rangle + |-1 \rangle \right ), \label{eq:|y>} \\ 
	& |z \rangle = |0 \rangle. \label{eq:|z>}
\end{align}  
Here, $| m\rangle$ can be represented in terms of the spherical harmonics of rank $1$, $Y_{l=1}^m (\theta ,\phi )$, with the polar angle $\theta$ against the $z$ axis and the azimuth angle $\phi$ against the $x$ axis. 
Thus $| \mu \rangle$ can also be expressed in terms of $Y_{l=1}^m (\theta ,\phi )$ as illustrated in Fig.~\ref{fig:1}. 

This paper is organized as follows. 
In Sec.~\ref{sec:3}, we derive the generalized Mermin-Ho relation for a spin-$1$ BEC and express it in terms of the su($3$) generators. 
In Sec.~\ref{sec:4}, we parametrize the su($3$) generators in terms of the polarization and the direction of the spin vector. 
We use this parametrization to construct spin-nematic vortices belonging to an su($2$) subalgebra of the su($3$) algebra. 
In Sec.~\ref{sec:5}, we derive a formula for the mass-current circulations, which is expressed in terms of four independent winding numbers, and apply it to ferromagnetic and polar-core vortices. 
In Sec. VI, we summarize the main results of this paper. 
The Gell-Mann matrices and the su($3$) structure constants are listed in Appendix~\ref{as1}, and the derivation of the original Mermin-Ho relation from the su($3$) Mermin-Ho relation is given in Appendix~\ref{as2}. 

\section{\label{sec:3}SU($3$) Mermin-Ho relation}
A mean-field state of a spin-$1$ BEC can be expanded in terms of the Cartesian basis $| \mu \rangle$ as 
\begin{align} 
	| \bm{\psi} (\bm{r} ,t) \rangle = \sum_{\mu} {\psi}_{\mu} (\bm{r} ,t) |\mu \rangle . \label{eq:mcwf} 
\end{align} 
Here, ${\psi}_{\mu} (\bm{r} ,t)$ is the $\mu$th component of the order parameter which gives the density as $\rho (\bm{r} ,t) \equiv \sum_{\mu} |{\psi}_{\mu} (\bm{r} ,t)|^2$. 
In the mean-field description, the spin and phase degrees of freedom are separable from the density degree of freedom; hence we can define the rescaled order parameter ${\xi}_{\mu} (\bm{r} ,t)$ as ${\xi}_{\mu} (\bm{r} ,t) \equiv {\psi}_{\mu} (\bm{r} ,t) / \sqrt{\rho (\bm{r} ,t)}$. 
In the following discussion, we omit the spatial and temporal coordinates $(\bm{r} ,t)$. 
The mass current can be expressed in terms of the rescaled order parameters as 
\begin{align} 
	\bm{v} \equiv \frac{\hbar}{2Mi} \sum_{\mu} [ {\xi}_{\mu}^* (\nabla {\xi}_{\mu}) - (\nabla {\xi}_{\mu}^*) {\xi}_{\mu}], \label{eq:v} 
\end{align} 
and its rotation can be transformed into 
\begin{align} 
	\nabla \times \bm{v} = \frac{i\hbar}{M} \sum_{\mu ,\nu ,\lambda} {\xi}_{\mu}^* {\xi}_{\nu} (\nabla {\xi}_{\nu}^* {\xi}_{\lambda} ) \times (\nabla {\xi}_{\lambda}^* {\xi}_{\mu} ), \label{eq:rotv1}
\end{align} 
where the suffixes $\mu$, $\nu$, and $\lambda$ over $x$, $y$, and $z$. 
Here, ${\xi}_{\mu}^* {\xi}_{\nu}$ in Eq.~(\ref{eq:rotv1}) can be interpreted as the transition amplitude from the state $\nu$ to the state $\mu$. 
This implies that Eq.~(\ref{eq:rotv1}) can be expressed in terms of the su($3$) roots, which connect two of the three basis states, or equivalently in terms of the Gel-Mann matrices ${\Lambda}_i$ ($i = 1, \cdots ,8$) (see Appendix A for their explicit representations)~\cite{Barnett,Yukawa}. 
In fact, as shown in Appendix~\ref{as1}, Eq.~(\ref{eq:rotv1}) can be rewritten as 
\begin{align} 
	\nabla \times \bm{v} = \frac{\hbar}{8M} \sum_{i,j,k=1}^8 f_{ijk} {\lambda}_i (\nabla {\lambda}_j ) \times (\nabla {\lambda}_k ), \label{eq:rotv2}
\end{align} 
where the su($3$) structure constant $f_{ijk}$ is given in Eq.~(\ref{eq:sf}) of Appendix~\ref{as1} and 
\begin{align} 
	{\lambda}_i \equiv \sum_{\mu ,\nu} ({\Lambda}_i)_{\mu \nu} {\xi}_{\mu}^* {\xi}_{\nu}, 
\end{align}  
which give the physical quantities such as the spin vector $f_{\mu}$ and observables concerning nematicity $q_{xy}$, $q_{yz}$, $q_{zx}$, $d_{x^2-y^2}$, and $d_{3z^2-f^2}$, as listed in Table~\ref{tab:1}. 
Here, the spin vector $f_{\mu}$ and the nematicity observables $q_{\mu \nu}$, $d_{x^2-y^2}$, and $d_{3z^2-f^2}$ are defined as follows: 
\begin{align} 
	&f_{\mu} = \sum_{\nu ,\lambda} (F_{\mu})_{\nu \lambda} {\xi}_{\nu}^* {\xi}_{\lambda}, \label{eq:f} \\ 
	&q_{\mu \nu} = \sum_{\lambda ,\eta} (F_{\mu} F_{\nu} + F_{\nu} F_{\mu} )_{\lambda \eta} {\xi}_{\lambda}^* {\xi}_{\eta}, \label{eq:q} \\ 
	&d_{x^2-y^2} = \sum_{\mu ,\nu} (F_x^2 - F_y^2 )_{\mu \nu} {\xi}_{\mu}^* {\xi}_{\nu}, \label{eq:d} \\ 
	&d_{3z^2-f^2} = \frac{1}{\sqrt{3}} \sum_{\mu ,\nu} (- F_x^2 - F_y^2 + 2 F_z^2)_{\mu \nu} {\xi}_{\mu}^* {\xi}_{\nu}. \label{eq:y}
\end{align} 
Here, $d_{3z^2-f^2}$ corresponds to the hyperchage. 
Equation~(\ref{eq:rotv2}) is the main result of this paper which we refer to as the su($3$) Mermin-Ho relation. 
It is a generalization of the Mermin-Ho relation to an arbitrary spin-$1$ BEC in which the magnitude of the spin can change over space and time. 
For a fully spin-polarized case, Eq.~(\ref{eq:rotv2}) reduces to the Mermin-Ho relation as shown in Appendix~\ref{as2}. 
\begin{table}[b]
\caption{\label{tab:1}Correspondence between the expectation values of the Gell-Mann matrices (upper row) and the observables concerning the spin vector and nematicity (lower row). }
\begin{ruledtabular}
\begin{tabular}{cccccccc}
${\lambda}_1$ & ${\lambda}_2$ & ${\lambda}_3$ & ${\lambda}_4$ & ${\lambda}_5$ & ${\lambda}_6$ & ${\lambda}_7$ & ${\lambda}_8$ \\ \colrule
$-q_{xy}$ & $f_z$ & $-d_{x^2-y^2}$ & $-q_{zx}$ & $-f_y$ & $-q_{yz}$ & $f_x$ & $d_{3z^2-f^2}$ 
\end{tabular}
\end{ruledtabular}
\end{table} 

\begin{figure*}[t]
	\begin{center} 
		\includegraphics[bb = 0 0 2197 452, clip, scale = 0.22]{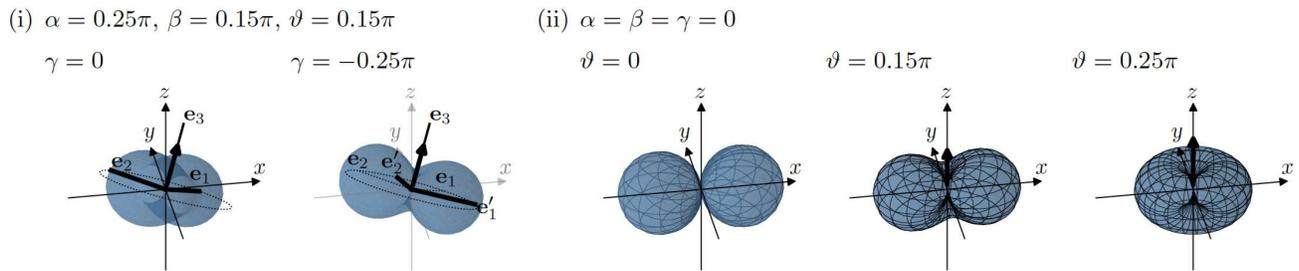}
	\end{center} 
	\caption{(Color Online) (i) Physical meanings of the parameters $\alpha$, $\beta$, and $\gamma$ giving the principal axes ${\bm{e}}_1$ and ${\bm{e}}_2$, and the normal direction ${\bm{e}}_3 = {\bm{e}}_1 \times {\bm{e}}_2$ through ${\bm{e}}_1 = (\cos {\alpha} \cos {\beta}, \sin {\alpha} \cos {\beta}, - \sin {\beta})^T$, ${\bm{e}}_2 = (- \sin {\alpha}, \cos {\alpha}, 0)^T$ and ${\bm{e}}_3 = (\cos {\alpha} \sin {\beta}, \sin {\alpha} \sin {\beta}, \cos {\beta})^T$. The parameters $\alpha$ and $\beta$ represent the azimuth and polar angles of the spin vector $\bm{f} = |\bm{f}| {\bm{e}}_3$ which is indicated by the thick solid arrow parallel to ${\bm{e}}_3$, and $\gamma$ is the rotation angle of the order parameter around ${\bm{e}}_3$. In the left panel, major and minor principal axes of the order parameter coincide with ${\bm{e}}_1$ and ${\bm{e}}_2$, respectively. In the right panel, the coordinate system is rotated about ${\bm{e}}_3$ through $\gamma$, where ${\bm{e}}_1^{\prime}$ and ${\bm{e}}_2^{\prime}$ give the major and minor principal axes in the rotated frame of reference. (ii) Parameter $\vartheta$ dependence of the order parameter, where $\mathcal{\theta}=0, 0.15\pi$, and $0.25\pi$ give the unpolarized (polar), partially polarized (broken-axisymmetry), and fully polarized (ferromagnetic) states, respectively. The wire frames are plotted as a guide to the eye. } 
	\label{fig:2}
\end{figure*} 
\section{\label{sec:4}Chechetkin-Anderson-Toulouse and Mermin-Ho vortices and their dual vortices \sout{in spin-1 systems}} 
\subsection{Chechetkin-Anderson-Toulouse and Mermin-Ho vortices} 
For later discussions, it is convenient to expressed $\xi_\mu$ in terms of five parameters $(\varphi ,\alpha ,\beta ,\gamma ,\vartheta )$ as follows: 
\begin{widetext} 
\begin{align} 
	\begin{pmatrix} {\xi}_x \\ {\xi}_y \\ {\xi}_z \end{pmatrix} &= e^{i\varphi} \exp {(-i\alpha F_z)} \exp {(-i\beta F_y)} \exp {(-i\gamma F_z)} \begin{pmatrix} \cos {\vartheta} \\ i \sin {\vartheta} \\ 0 \end{pmatrix} \nonumber \\
	&= e^{i\varphi} \begin{pmatrix} 
	(\cos {\alpha} \cos {\beta} \cos {\gamma} - \sin {\alpha} \sin {\gamma} ) \cos {\vartheta} - i (\cos {\alpha} \cos {\beta} \sin {\gamma} + \sin {\alpha} \cos {\gamma} ) \sin {\vartheta} \\ 
	(\sin {\alpha} \cos {\beta} \cos {\gamma} + \cos {\alpha} \sin {\gamma} ) \cos {\vartheta} + i (- \sin {\alpha} \cos {\beta} \sin {\gamma} + \cos {\alpha} \cos {\gamma} ) \sin {\vartheta} \\ 
	- \sin {\beta} \cos {\gamma} \cos {\vartheta} + i \sin {\beta} \sin {\gamma} \sin {\vartheta} 
	\end{pmatrix}. \label{eq:wf}
\end{align} 
\end{widetext} 
Here, the parameter $\varphi$ is the U($1$) phase, $\alpha$ and $\beta$ represent the azimuth and polar angles of the spin vector $\bm{f}$ in the Cartesian coordinates, and $\gamma$ is the rotation angle about the direction of $\bm{f}$ as illustrated in Figs.~\ref{fig:2} (i). 
The parameter $\vartheta$ indicates the polarization of the spin vector as $|\bm{f}| = |\sin {2\vartheta}|$. 
Thus, $\vartheta = 0$, $\pi / 2$, $\pi$, $\cdots$ show the unpolarized states, $\vartheta = \pi / 4$, $3\pi / 4$, $\cdots$ show the fully polarized states, and $\vartheta \in (0, \pi / 4)$, $(\pi / 4, \pi / 2)$, $\cdots$ show partially polarized states such as broken-axisymmetry states~\cite{Murata} (see Figs.~\ref{fig:2} (ii)). 
Then, the spin vector and the nematicity observables in Eqs.~(\ref{eq:f})-(\ref{eq:y}) can be expressed in terms of $\alpha$, $\beta$, $\gamma$, and $\vartheta$ as 
\begin{widetext} 
\begin{align} 
	&\bm{f} = \sin {2\vartheta} \begin{pmatrix} \cos {\alpha} \sin {\beta} \\ \sin {\alpha} \sin {\beta} \\ \cos {\beta} \end{pmatrix}, \label{eq:f_param} \\ 
	&q_{xy} = \frac{1}{2} \{ \sin {2\alpha} \ {\sin}^2 {\beta} - \cos {2\vartheta} [ \sin {2\alpha} (1 + {\cos}^2 {\beta} ) \cos {2\gamma} + 2 \cos {2\alpha} \cos {\beta} \sin {2\gamma} ] \}, \label{eq:qxy_param} \\ 
	&q_{yz} = \sin {\beta} [ \sin {\alpha} \cos {\beta} + \cos {2\vartheta} ( \sin {\alpha} \cos {\beta} \cos {2\gamma} + \cos {\alpha} \sin {2\gamma} ) ], \label{eq:qyz_param} \\ 
	&q_{zx} = \sin {\beta} [ \cos {\alpha} \cos {\beta} + \cos {2\vartheta} (\cos {\alpha} \cos {\beta} \cos {2\gamma} - \sin {\alpha} \sin {2\gamma} ) ], \label{eq:qzx_param} \\  
	&d_{x^2-y^2} = \frac{1}{2} \{ \cos {2\alpha} \ {\sin}^2 {\beta} - \cos {2\vartheta} [ \cos {2\alpha} ( 1 + {\cos}^2 {\beta} ) \cos {2\gamma} - 2 \sin {2\alpha} \cos {\beta} \sin {2\gamma} ] \}, \label{eq:dxy_param} \\ 
	&d_{3z^2-f^2} = \frac{1}{2\sqrt{3}} [ - 1 + 3 {\cos}^2 {\beta} ( 1 - \cos {2\vartheta} \cos {2\gamma} ) ]. \label{eq:y_param}
\end{align} 
\end{widetext} 

\begin{figure*}[t]
	\begin{center} 
		\includegraphics[bb = 0 0 2058 831, clip, scale = 0.24]{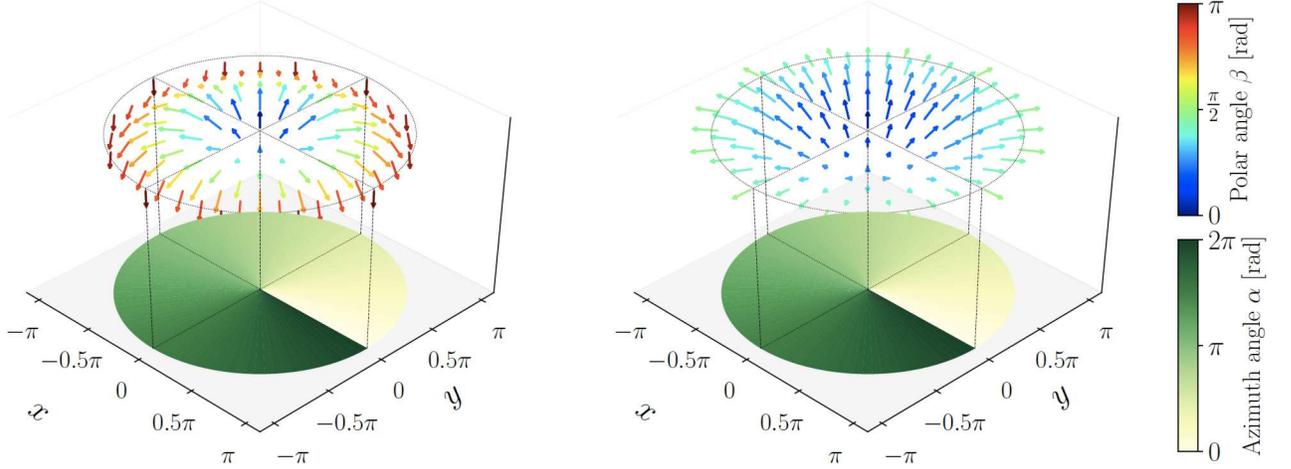}
	\end{center} 
	\caption{(Color Online) Fully polarized spin textures in CAT (i) and MH (ii) vortices. Each arrow represents the direction of a local spin vector with the color showing the value of $\beta$ according to the upper right gauge. The projection circle below represents the value of $\alpha$ according to the lower right gauge. In both (i) and (ii), $\alpha$ changes by $2\pi$ along the circumference of the vortex, while $\beta$ changes by $\pi$ in (i) and $\pi/2$ in (ii) in the radial direction. }  
	\label{fig:3}
\end{figure*}

When the spin is fully polarized, which corresponds, for example, to $\vartheta = \pi / 4$, Eqs.~(\ref{eq:f_param})-(\ref{eq:y_param}) imply that the nematicity observables can be expressed in terms of the spin vector as $q_{\mu \nu} = f_{\mu} f_{\nu}$, $d_{x^2-y^2} = (f_x^2 - f_y^2) / 2$, and $d_{3z^2-f^2} = (- f_x^2 - f_y^2 + 2 f_z^2) / 2\sqrt{3}$, and Eq.~(\ref{eq:rotv2}) reduces to be the following Mermin-Ho relation~\cite{Mermin2}: 
\begin{align} 
	\nabla \times \bm{v} = \frac{\hbar}{2M} {\epsilon}_{\mu \nu \lambda} f_{\mu} (\nabla f_{\nu} ) \times (\nabla f_{\lambda} ). \label{eq:MH}
\end{align}  
Here, the completely antisymmetric tensor ${\epsilon}_{\mu \nu \lambda}$ is nothing but the so($3$) structure constant. 
The derivation of Eq.~(\ref{eq:MH}) from Eq.~(\ref{eq:rotv2}) is shown in Appendix~\ref{as2}. 
The Mermin-Ho relation can describe the well-known Chechetkin-Anderson-Toulouse (CAT) and Mermin-Ho (MH) vortices shown in Fig.~\ref{fig:3} and we can obtain their winding numbers from this relation. 

\subsection{Dual Chechetkin-Anderson-Toulouse and Mermin-Ho vortices}
\begin{figure*}[t]
	\begin{center} 
		\includegraphics[bb = 0 0 2060 1751, clip, scale = 0.24]{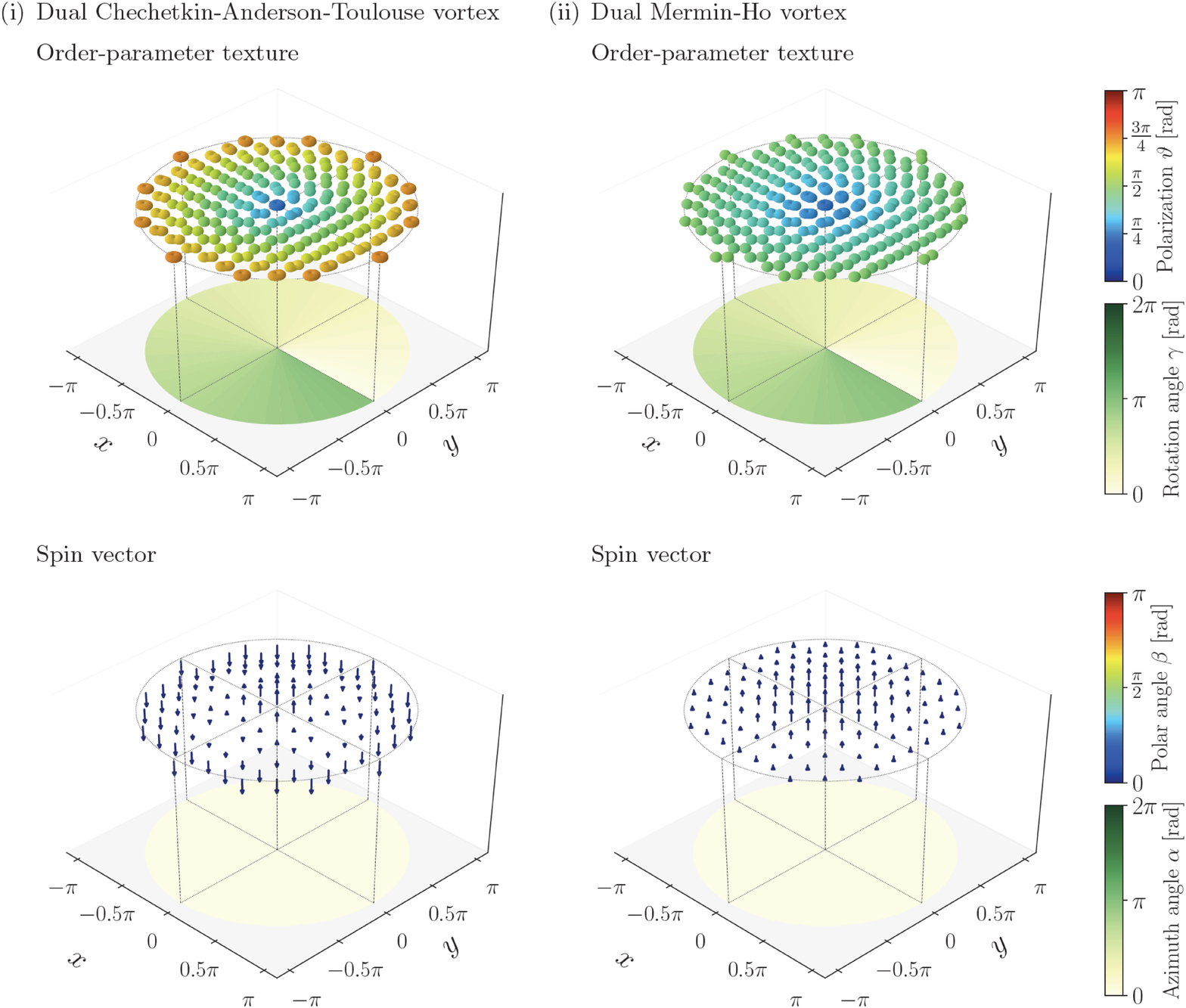}
	\end{center} 
	\caption{(Color Online) Spatial distributions of the order parameter (first row) and the spin vector (second row) in spin-nematic vortices which are dual to the Chechetkin-Anderson-Toulouse and Mermin-Ho vortices in Fig.~\ref{fig:3}. In the upper panels, the color of the order parameter indicates the value of $\vartheta$ according to the upper right gauge and the color on the projection circle of the wave-function texture represents the value of $\gamma$ according to the lower right gauge. In the lower panels, the color maps of the spin texture and its projection circle are the same as those in Fig.~\ref{fig:3}. (i) In a vortex dual to the Chechetkin-Anderson-Toulouse vortex,  $\gamma$ and $\vartheta$ change by $\pi$ and $\pi / 2$. In terms of the pseudo-spin vector ${\bm{f}}^{\prime}$, ${\alpha}^{\prime}$ and ${\beta}^{\prime}$ change by $2\pi$ and $\pi$, as in $\alpha$ and $\beta$ of the CAT vortex. (ii) In a vortex dual to the Mermin-Ho vortex, $\gamma$ and $\vartheta$ change by $\pi$ and $\pi / 4$, which implies that ${\alpha}^{\prime}$ and ${\beta}^{\prime}$ change by $2\pi$ and $\pi / 2$. On the other hand, $\alpha$ and $\beta$ stay constant in both cases. } 
	\label{fig:4}
\end{figure*}

In spin-$1$ BECs, the spin nematicity can also form a texture in a vortex and we can consider such spin-nematic vortices analogous to the CAT and MH vortices. 
We shall refer to such vortices as dual CAT and MH vortices.
When $\alpha = \beta = 0$, Eqs.~(\ref{eq:f_param})-(\ref{eq:y_param}) reduce to 
\begin{align} 
	&f_x = f_y = 0, \ f_z = \sin {2\vartheta}, \label{eq:sn_fz} \\ 
	&q_{xy} = - \sin {2\gamma} \cos {2\vartheta}, \ q_{yz} = q_{zx} = 0, \label{eq:sn_qxy} \\ 
	&d_{x^2-y^2} = - \cos {2\gamma} \cos {2\vartheta}, \label{eq:sn_dxy} \\ 
	&d_{3z^2-f^2} = \frac{1}{\sqrt{3}}.  
\end{align} 
In terms of new parameters $2 \gamma \equiv {\alpha}^{\prime}$ and $2 \vartheta - \pi / 2 \equiv {\beta}^{\prime}$, the spin-vector and nematic-tensor quantities together can be cast into the form of a unit-length pseudo-spin ${\bm{f}}^{\prime} \equiv (d_{x^2-y^2} ,q_{xy} ,f_z)^T = (\cos {{\alpha}^{\prime}} \sin {{\beta}^{\prime}} , \sin {{\alpha}^{\prime}} \sin {{\beta}^{\prime}} , \cos {{\beta}^{\prime}})^T$. 
Then, we can construct vortices dual to the CAT and MH vortices as shown in Fig.~\ref{fig:4}. 
Among these two types of vortices, the spin-nematic vortex shown in Fig.~\ref{fig:4} (i) is identified as the ferromagnetic core vortex in Figs.~6 (c) and (d) in Ref.~\cite{SKobayashi}. 
Here, we note that $f_z$, $q_{xy}$, and $d_{x^2-y^2}$ in the pseudo-spin vector ${\bm{f}}^{\prime}$ form an su($2$) subalgebra of the su($3$), which is characterized by the structure constant $f_{123} = 2$. 
The Mermin-Ho relation in Eq.~(\ref{eq:rotv2}) for these vortices can be expressed in terms of the pseudo-spin ${\bm{f}}^{\prime}$ as  
\begin{align} 
	\nabla \times \bm{v} = \frac{\hbar}{4M} {\epsilon}_{\mu \nu \lambda} f_{\mu}^{\prime} (\nabla f_{\nu}^{\prime} ) \times (\nabla f_{\lambda}^{\prime} ). \label{eq:MH_prime}
\end{align} 
The right-hand side of Eq.~(\ref{eq:MH_prime}) is half of the original Mermin-Ho relation in Eq.~(\ref{eq:MH}). 

\begin{figure}[t] 
	\begin{center} 
		\includegraphics[bb = 0 0 800 600, clip, scale = 0.25]{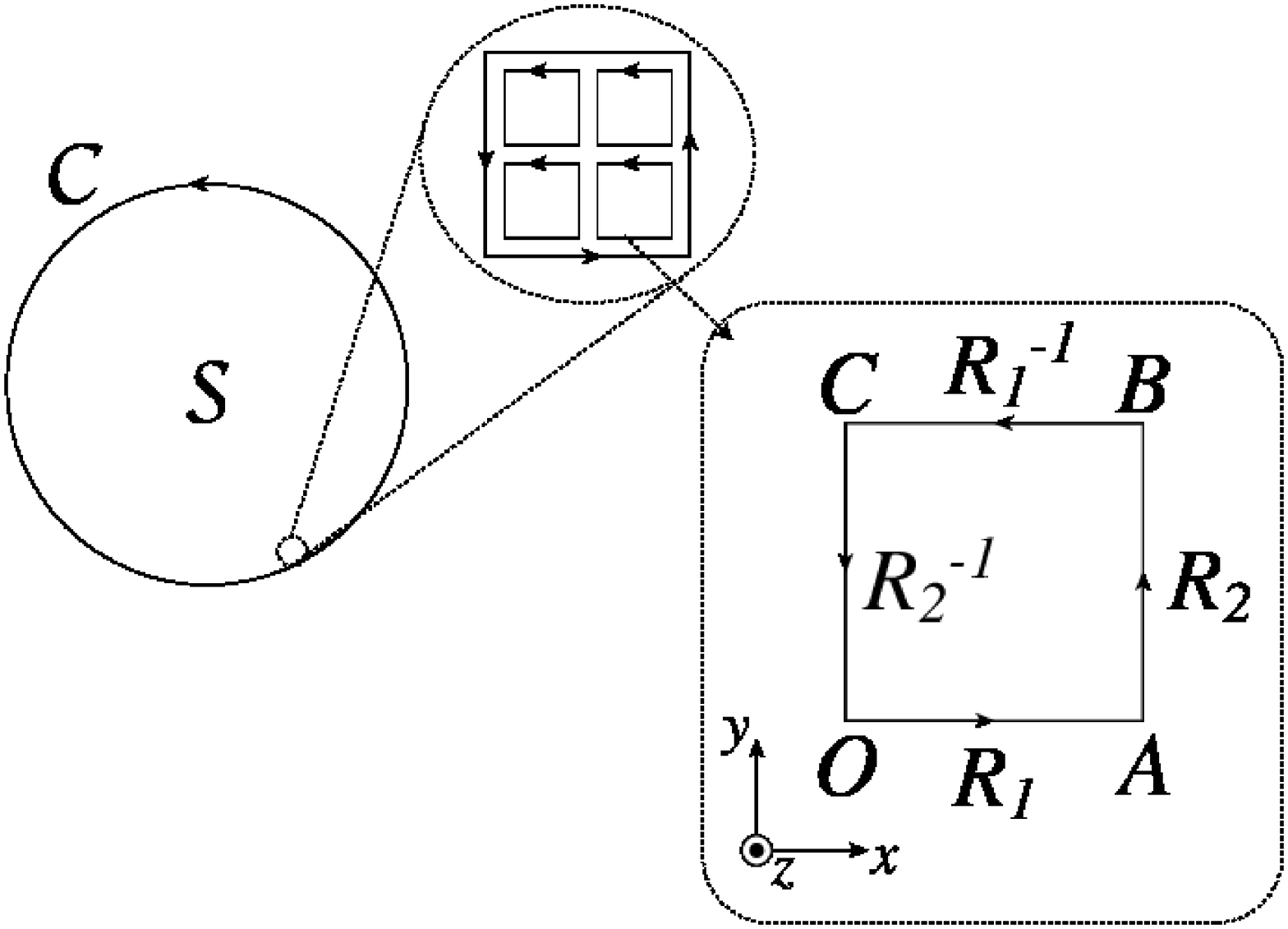} 
	\end{center} 
	\caption{Circumference of a vortex $\mathcal{C}$ (left) decomposed into infinitesimal rectangular loops (upper middle), one of which is enlarged and labeled as $OABC$ (right) with $R_1$ and $R_2$ showing SO($3$) rotations given in Eqs.~(\ref{eq:R1}) and (\ref{eq:R2}), where $\mathcal{S}$ indicates the area enclosed by $\mathcal{C}$. }
	\label{fig:5}
\end{figure} 
\subsection{Nonholonomy of dual CAT and MH vortices} 
We follow Ref.~\cite{Leggett} to analyze the Mermin-Ho relation for the spin-nematic vortices in Eq.~(\ref{eq:MH_prime}) from a viewpoint of nonholonomy. 
Suppose that $\mathcal{C}$ is a two-dimensional circle enclosing a spin-nematic vortex at the center $\mathcal{C}$ of Fig.~\ref{fig:5}. 
The phase that the pseudo-spin vector ${\bm{f}}^{\prime}$ acquires along the circumference of $\mathcal{C}$ can be obtained as follows. 
First, we decompose the circle into infinitesimal rectangular loops $O \to A \to B\to C \to O$ as shown in Fig.~\ref{fig:5}. 
Along this loop, we consider two non-commutative SO($3$) rotations 
\begin{align} 
	& R_1 = \exp {(- i \delta {\phi}_1 {\bm{n}}_1 \cdot {\bm{F}}^{\prime})}, \label{eq:R1} \\ 
	& R_2 = \exp {(- i \delta {\phi}_2 {\bm{n}}_2 \cdot {\bm{F}}^{\prime})}, \label{eq:R2} 
\end{align} 
where the unit vector $\bm{n}_i$ and the angle $\delta {\phi}_i$ ($i=1,2$) represent the axis and the angle of the rotation, and the components of ${\bm{F}}^{\prime}$ are given by $F^{\prime}_x \equiv D_{x^2-y^2} / 2= - {\Lambda}_3 / 2$, $F^{\prime}_y \equiv Q_{xy} / 2= - {\Lambda}_1 / 2$, and $F^{\prime}_z \equiv F_z / 2= {\Lambda}_2 / 2$. 
The factor $2$ in the denominator in the definition of $F_{\mu}^{\prime}$ is due to the structure constant $f_{123} = 2$ among ${\Lambda}_i$ ($i = 1,2,3$). 
On the paths $O \to A$, $A \to B$, $B \to C$, and $C \to O$, we apply $R_1$, $R_2$, $R_1^{-1}$, and $R_2^{-1}$, respectively, to ${\bm{f}}^{\prime}$; then the total action on ${\bm{f}}^{\prime}$ is given up to the second order of $\delta {\phi}_i$ as 
\begin{align} 
	R_2^{-1} R_1^{-1} R_2 R_1 = & I_3 + \delta {\phi}_1 \delta {\phi}_2 ({\bm{n}}_1 \times {\bm{n}}_2 ) \cdot {\bm{F}}^{\prime} \nonumber \\ 
	&+ \mathcal{O} (\delta {\phi}_i^3), \label{eq:operator1}
\end{align} 
where $I_3$ indicates the three-dimensional identity matrix.  
The infinitesimally small vectors $\delta {\phi}_1 {\bm{n}}_1$ and $\delta {\phi}_2 {\bm{n}}_2$ can be expressed in terms of ${\bm{f}}^{\prime}$ and its spatial derivative as 
\begin{align} 
	\delta {\phi}_1 {\bm{n}}_1 = \delta x {\bm{f}}^{\prime} \times ({\nabla}_x {\bm{f}}^{\prime} ), \ \delta {\phi}_2 {\bm{n}}_2 = \delta y {\bm{f}}^{\prime} \times ({\nabla}_y {\bm{f}}^{\prime} ), 
\end{align} 
where $\delta x$ and $\delta y$ are the lengths between $O$ and $A$ and between $B$ and $C$. 
Then, Eq.~(\ref{eq:operator1}) can be expressed as 
\begin{align} 
	R_2^{-1} R_1^{-1} R_2 R_1 \simeq I_3 + \delta x \delta y {\epsilon}_{\mu \nu \lambda} f_{\mu}^{\prime} ({\nabla}_x f_{\nu}^{\prime} ) ({\nabla}_y f_{\lambda}^{\prime}) ({\bm{f}}^{\prime} \cdot {\bm{F}}^{\prime}). \label{eq:operator2}
\end{align} 
In the limit of $\delta x \to 0$ and $\delta y \to 0$, the second term on the right-hand side of Eq.~(\ref{eq:operator2}) can be considered as the generator of the phase ${\bm{f}}^{\prime}$ and the gained phase can be expressed as 
\begin{align} 
	\frac{1}{2} \int_{\square} d\bm{S} \cdot [{\epsilon}_{\mu \nu \lambda} f_{\mu}^{\prime} (\nabla f_{\nu}^{\prime} ) \times (\nabla f_{\lambda}^{\prime} )],  
\end{align} 
where $\square$ represents the rectangle $OABC$. 
On the other hand, $\delta \chi$ can be expressed in terms of the local phase $\chi$ as 
\begin{align} 
	\delta \chi = \oint_{\square} d\bm{l} \cdot (\nabla \chi ), 
\end{align}  
which implies 
\begin{align} 
	\oint_{\square} d\bm{l} \cdot (\nabla \chi ) = \frac{1}{2} \int_{\square} d\bm{S} \cdot [{\epsilon}_{\mu \nu \lambda} f_{\mu}^{\prime} (\nabla f_{\nu}^{\prime} ) \times (\nabla f_{\lambda}^{\prime} )]. 
\end{align} 
Summing up all small loops, we can obtain the total phase $\Delta \chi$ that ${\bm{f}}^{\prime}$ gains along $\mathcal{C}$ as 
\begin{align}
	\Delta \chi = \oint_{\mathcal{C}} d\bm{l} \cdot (\nabla \chi ) = \frac{1}{2} \int_{\mathcal{S}} d\bm{S} \cdot [{\epsilon}_{\mu \nu \lambda} f_{\mu}^{\prime} (\nabla f_{\nu}^{\prime} ) \times (\nabla f_{\lambda}^{\prime} )], 
\end{align} 
where $\mathcal{S}$ indicates the area enclosed by $\mathcal{C}$. 
The Mermin-Ho relation for spin vortices and that for spin-nematic vortices in Eqs.~(\ref{eq:MH}) and (\ref{eq:MH_prime}) imply that the circulation of the mass current around a vortex is given by 
\begin{align} 
	\oint_{\mathcal{C}} d\bm{l} \cdot \bm{v} = \frac{\hbar}{M} \left ( \oint_{\mathcal{C}} d\bm{l} \cdot (\nabla \varphi ) + \frac{1}{2} \Delta \chi \right ). \label{eq:vcirc}
\end{align} 
The first term on the right-hand side of Eq.~(\ref{eq:vcirc}) is nonvanishing when the U($1$) phase is singular at the center of a vortex. 
As discussed later in Eq.~(\ref{eq:singularity}), the spin-nematic vortices dual to the CAT and MH vortices have singularities at the center of the vortex, so $\oint_{\mathcal{C}} d\bm{l} \cdot (\nabla \varphi ) = \pi$ in both cases. 
The second term on the right-hand side of Eq.~(\ref{eq:vcirc}) indicates the nonholonomy of the spinor order parameter. 
The phases are given by $\Delta \chi = 2\pi$ and $\pi$, which are the same as the cases of the ordinary CAT and MH vortices~\cite{Leggett}; however, due to the difference in the structure constant, the circulations of the mass currents are given by $3h/2M$ for the dual CAT vortex and $h/M$ for the dual MH vortex. 

\section{\label{sec:5}Winding number of a spin-nematic texture}
Next, we examine the circulation of the mass current $\bm{v}$. 
Here, $\bm{v}$ can be expressed in terms of the set of parameters introduced in the preceding section as 
\begin{align} 
	\bm{v} = \frac{\hbar}{M} \{ (\nabla \varphi ) - [ (\nabla \alpha ) \cos {\beta} + (\nabla \gamma ) ] \sin {2\vartheta} \}. \label{eq:mc}
\end{align} 
Equation (35) reduces to $\bm{v} = (\hbar / M) \{ [ \nabla (\varphi \mp \gamma ) ] \mp (\nabla \alpha ) \cos {\beta} \}$ ($-(+)$ sign for $\vartheta = \pm \pi / 4$ ($\pm 3\pi / 4$)) for the fully-polarized case and $\bm{v} = (\hbar / M) (\nabla \varphi )$ for the unpolarized case. 
When the spin polarization is constant, the circulation of $\bm{v}$ around a vortex core can be computed on the basis of these expressions. 

In general, however, the spin polarization can change over space and time and form a spin-nematic texture. 
To derive the expression of the mass-current circulation around a vortex with nonuniform polarization, we consider a situation in which spin-$1$ bosons are confined in a two-dimensional disk with unit radius in the $x$-$y$ plane and the mass current $\bm{v}$ flows in the $x$-$y$ plane. 
The core of the vortex, which is assumed to locate at the center of the disk, should be fully polarized along the $+z$ or $-z$ direction, i.e., $\bm{\xi} \propto (1, i, 0)^{\mathrm{T}}$ or $(1, -i, 0)^{\mathrm{T}}$ ($| \pm 1\rangle$ in the irreducible representation), or unpolarized, i.e., $\bm{\xi} \propto (0, 0, 1)^{\mathrm{T}}$ ($|0 \rangle$ in the irreducible representation) due to the symmetry around the vortex core.  
Setting the cylindrical coordinate $(R, \Phi )$ with the vortex core at the origin, we display the spatial dependence of the five parameters as $\alpha (R, \Phi )$, $\beta (R, \Phi )$, $\gamma (R, \Phi )$, and $\vartheta (R, \Phi )$. 
We assume 
\begin{align} 
	& \alpha (R, \Phi ) = n_{\alpha} \Phi , \label{eq:alpha} \\ 
	& \beta (R, \Phi ) = \frac{\pi}{2} ( {\tilde{\beta}}_0 + n_{\beta} R) , \label{eq:beta} \\ 
	& \gamma (R, \Phi ) = \frac{1}{2} n_{\gamma} \Phi , \label{eq:gamma} \\ 
	& \vartheta (R, \Phi ) = \frac{\pi}{4} ( {\tilde{\vartheta}}_0 + n_{\vartheta} R ), \label{eq:theta}
\end{align} 
where ${\tilde{\beta}}_0$ and ${\tilde{\vartheta}}_0$ depend on the state of the vortex core as noted in Table~\ref{tab:2}, and the winding numbers $n_{\alpha}$, $n_{\beta}$, $n_{\gamma}$, and $n_{\vartheta}$ are determined by the boundary conditions on the circumference of the disk. 
It follows from Table~\ref{tab:2} that the ferromagnetic and polar-core states are respectively given by $e^{i(\varphi - \alpha - \gamma)} (1, i, 0)^T / \sqrt{2}$ and $e^{i\varphi} (-\sin {\alpha} \sin {\gamma} , \cos {\alpha} \sin {\gamma} , \cos {\gamma} )^T$, and a vortex with a ferromagnetic core can always be filled, whereas a vortex with a polar core can be filled only when $n_{\gamma} = 0$. 
In Eqs.~(\ref{eq:alpha}) and (\ref{eq:gamma}), we assume that $\alpha (R, \Phi = 0 ) = 0$ and $\gamma (R, \Phi = 0) = 0$ without loss of generality, since a vortex with $\alpha (R, 0 ) \neq 0$ or $\gamma (R, 0) \neq 0$ can be obtained from the vortex in Eqs.~(\ref{eq:alpha}) and (\ref{eq:gamma}) by a uniform rotation of the spin vector or the orientation of the order parameter, leaving the winding number unchanged.    

Let us consider the mass-current circulation for vortices with the boundary condition given in Eqs.~(\ref{eq:alpha})-(\ref{eq:theta}).  
Substituting the boundary conditions into Eq.~(\ref{eq:mc}), we obtain the circulation of $\bm{v}$ as  
\begin{widetext} 
\begin{align}
	\oint {\bm{v}} \cdot d\bm{l} = \oint (\nabla \varphi ) \cdot d \bm{l} + \frac{h}{M} \biggl \{ & n_{\alpha} \left [ \cos {\frac{\pi}{2} {\tilde{\beta}}_0} \sin {\frac{\pi}{2} {\tilde{\vartheta}}_0} - \cos {\frac{\pi}{2} ( {\tilde{\beta}}_0 + n_{\beta} )} \sin {\frac{\pi}{2} ( {\tilde{\vartheta}}_0 + n_{\vartheta})} \right ] \nonumber \\ 
	+ & \frac{1}{2} n_{\gamma} \left [ \sin {\frac{\pi}{2} {\tilde{\vartheta}}_0} - \sin {\frac{\pi}{2} ( {\tilde{\vartheta}}_0 + n_{\vartheta} )} \right ] \biggr \} , \label{eq:circv}
\end{align} 
\end{widetext} 
where the first term on the rignt-hand side is determined to satisfy the single-valuedness of the order parameter. 
\begin{table}[b]
\caption{\label{tab:2}Boundary conditions at the center of a vortex with a ferromagnetic core and that with a polar core.}
\begin{ruledtabular}
\begin{tabular}{ccc}
& Ferromagnetic core & Polar core \\ 
\colrule \addlinespace[2pt]
${\tilde{\beta}}_0$ & $0$ & $-1$ \\ 
${\tilde{\vartheta}}_0$ & $1$ & $0$ \\ 
\end{tabular}
\end{ruledtabular}
\end{table}

Let us apply Eq.~(\ref{eq:circv}) to vortices with ferromagnetic cores of the CAT and MH vortices in Fig.~\ref{fig:3} and their spin-nematic version in Fig.~\ref{fig:4}. 
In this case, the circulation of the mass current can be obtained from Eq.~(\ref{eq:circv}) and Table~\ref{tab:2} as 
\begin{widetext} 
\begin{align} 
	&\oint {\bm{v}} \cdot d\bm{l} = \frac{\hbar}{M} \oint (\nabla \varphi ) \cdot d \bm{l} + \frac{h}{M} \left [ n_{\alpha} \left ( 1 - \cos {\frac{\pi}{2} n_{\beta}} \cos {\frac{\pi}{2} n_{\vartheta}} \right ) 
	+ \frac{1}{2} n_{\gamma} \left ( 1 - \cos {\frac{\pi}{2} n_{\vartheta}} \right ) \right ]. \label{eq:circv_f} 
\end{align} 
\end{widetext} 
The winding numbers characterizing the CAT and MH vortices are given by $(n_{\alpha} ,n_{\beta} ,n_{\gamma} ,n_{\vartheta} ) = (1, 2, 0, 0)$ and $(1, 1, 0, 0)$, respectively, and their U($1$) phases $\varphi$ are not singular at the core. 
Hence their circulations are given by 
\begin{align} 
	\oint {\bm{v}} \cdot d\bm{l} 
	= \left \{ \begin{array}{ll}  
	\frac{2h}{M} &(\text{CAT}); \\ 
	\frac{h}{M} & (\text{MH}). 
	\end{array} \right . 
\end{align} 

\begin{figure*}[t]
	\begin{center} 
		\includegraphics[bb = 0 0 2058 1703, clip, scale = 0.23]{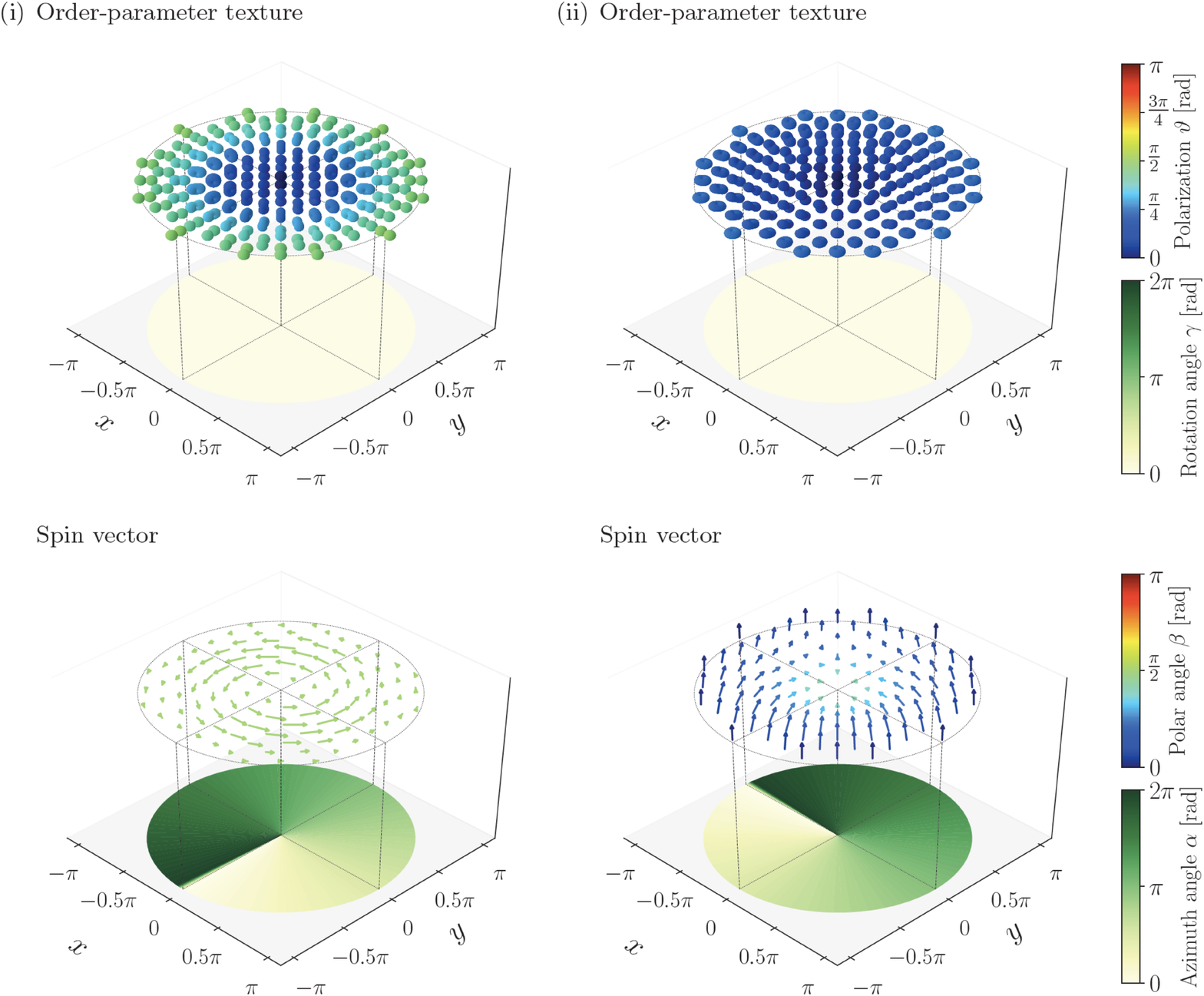}
	\end{center} 
	\caption{(Color Online) Order-parameter and spin-vector textures in the polar-core vortices in Fig.~1 (b) of Ref.~\cite{Kawaguchi} and Fig.~6 (b) and (d) in Ref.~\cite{SKobayashi} displayed in the same manner as in Fig.~\ref{fig:4}. Figures (i) and (ii) correspond to the vortices in Refs.~\cite{Kawaguchi} and~\cite{SKobayashi}, respectively. } 
	\label{fig:6}
\end{figure*} 
On the other hand, the spin-nematic vortices dual to the CAT and MH vortices have $(n_{\alpha} ,n_{\beta} ,n_{\gamma} ,n_{\vartheta} ) = (0, 0, 1, 2)$ and $(0, 0, 1, 1)$, respectively, and the U($1$) phase $\varphi$ should increase by $\pi$ along circumference of the vortex for both cases. 
This can be confirmed by substituting Eqs.~(\ref{eq:gamma}) and (\ref{eq:theta}) into the rescaled order parameter in Eq.~(\ref{eq:wf}) with $\alpha = \beta = 0$ giving 
\begin{align} 
	\bm{\xi} = \frac{e^{i\varphi (R, \Phi )}}{\sqrt{2}} \begin{pmatrix} 
	e^{- \frac{i}{2} \Phi} \cos {\frac{\pi}{4} n_{\vartheta} R} - e^{\frac{i}{2} \Phi} \sin {\frac{\pi}{4} n_{\vartheta} R} \\ 
	i (e^{- \frac{i}{2} \Phi} \cos {\frac{\pi}{4} n_{\vartheta} R} + e^{\frac{i}{2} \Phi} \sin {\frac{\pi}{4} n_{\vartheta} R}) \\ 0
	\end{pmatrix}. \label{eq:wf_sn}
\end{align} 
Equation~(\ref{eq:wf_sn}) should be single-valued at $\Phi = 0$ and $2\pi$, that is to say,  
\begin{align} 
	&\bm{\xi} (R, 0) = \frac{e^{i\varphi (R, 0)}}{\sqrt{2}} \begin{pmatrix} 
	\cos {\frac{\pi}{4} n_{\vartheta} R} - \sin {\frac{\pi}{4} n_{\vartheta} R} \\ 
	i (\cos {\frac{\pi}{4} n_{\vartheta} R} + \sin {\frac{\pi}{4} n_{\vartheta} R}) \\ 0 \end{pmatrix} \nonumber \\ 
	= \ & \bm{\xi} (R, 2\pi ) = - \frac{e^{i\varphi (R, 2\pi )}}{\sqrt{2}} \begin{pmatrix} 
	\cos {\frac{\pi}{4} n_{\vartheta} R} - \sin {\frac{\pi}{4} n_{\vartheta} R} \\ 
	i (\cos {\frac{\pi}{4} n_{\vartheta} R} + \sin {\frac{\pi}{4} n_{\vartheta} R}) \\ 0 \end{pmatrix}, \label{eq:singularity}
\end{align} 
which implies that $\varphi (R, 2\pi ) - \varphi (R, 0) = \pi$. 
In this case, however, the core of a vortex is filled because it is ferromagnetic and has the spin-gauge symmetry, which implies that the U($1$) phase at the center is always given by ${\varphi}^{\prime} (0, \Phi ) \equiv \varphi (0, \Phi ) - \gamma (0, \Phi )= 0$. 
Thus, the circulations for the spin-nematic vortices are given by  
\begin{align} 
	\oint {\bm{v}} \cdot d\bm{l} 
	= \left \{ \begin{array}{ll}  
	\frac{3h}{2M} & (\text{dual CAT}); \\ 
	\frac{h}{M} & (\text{dual MH}). 
	\end{array} \right . 
\end{align} 
The mass-current circulation around the dual CAT vortex takes a half-quantized value unlike that around the CAT vortex. 
On the other hand, the mass-current circulation around the dual MH vortex coincides with that of the MH vortex, in spite of the differences in the structure constants and the U($1$) phase. 

The circulation of the mass current around a polar-core vortex can also be calculated from Eq.~(\ref{eq:circv}). 
In this case, ${\tilde{\beta}}_0 = -1$ and ${\tilde{\vartheta}}_0 = 0$ according to Table~\ref{tab:2}, and Eq.~(\ref{eq:circv}) becomes 
\begin{widetext} 
\begin{align}
	\oint {\bm{v}} \cdot d\bm{l} = \frac{\hbar}{M} \oint (\nabla \varphi ) \cdot d \bm{l} + \frac{h}{M} \left ( n_{\alpha} \sin {\frac{\pi}{2} n_{\beta}} - \frac{1}{2} n_{\gamma} \right ) \sin {\frac{\pi}{2} n_{\vartheta}}. \label{eq:circv_p} 
\end{align} 
\end{widetext} 
Let us take the polar-core vortices in Refs.~\cite{Kawaguchi,SKobayashi} as an example. 
The polar-core vortex in (b) of Fig.~1 in Ref.~\cite{Kawaguchi} can be characterized by the spin vector lying in the $xy$ plane whose azimuth angle $\alpha$ changes by $2\pi$ along the circumference. 
The polarization $\vartheta$ changes by $\pi / 2$ along the radius direction. 
On the other hand, the polarization parameter $\vartheta$ and the polar angle $\beta$ of the polar-core vortex in (b) and (d) of Fig.~6 in Ref.~\cite{SKobayashi} changes by $\pi / 4$ and $\pi / 2$ in the radius and the azimuth angle $\alpha$ changes by $2\pi$ along the circumference of the vortex. 
The spin and wave-function textures of these vortices are illustrated in Figs.~\ref{fig:6}. 
Their winding numbers are given by $(n_{\alpha} ,n_{\beta} ,n_{\gamma} ,n_{\vartheta}) = (1, 0, 0, 2)$ and $(1, 1, 0, 1)$. 
The U($1$) phases should satisfy $\varphi (R, 2\pi) - \varphi (R, 0) = 0$ for the cases in Refs.~\cite{Kawaguchi} and~\cite{SKobayashi}. 
Then, the circulations of the mass currents can be obtained as 
\begin{align} 
	\oint {\bm{v}} \cdot d\bm{l} 
	= \left \{ \begin{array}{ll}  
	0 & (\text{Ref.~\cite{Kawaguchi}}); \\ 
	\frac{h}{M} & (\text{Ref.~\cite{SKobayashi}}). 
	\end{array} \right. 
\end{align} 
Here, we note that the texture of the polar-core vortex in Ref.~\cite{Kawaguchi} is obtained from the vortex with the boundary conditions in Eqs.~(\ref{eq:alpha})-(\ref{eq:theta}) by a uniform $-\pi /2$ spin-rotation $\exp {(i \pi F_z / 2)}$ around the $z$ axis, which does not affect the mass-current circulation. 

\begin{figure}[t]
	\begin{center} 
		\includegraphics[bb = 0 0 1024 727, clip, scale = 0.23]{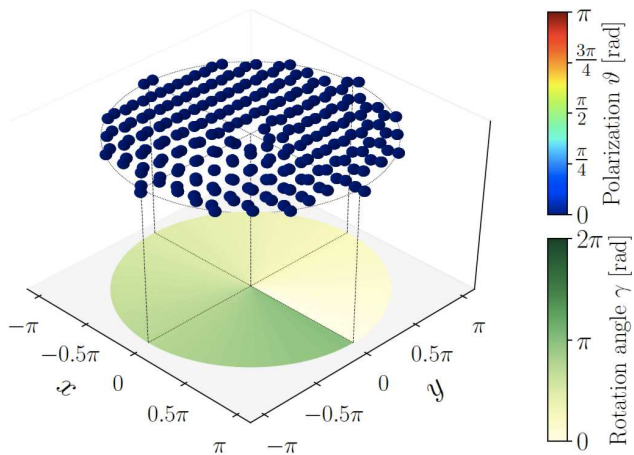}
	\end{center} 
	\caption{(Color Online) Order-parameter texture of the polar-core half-quantum vortex characterized by the winding number of $(n_{\alpha} ,n_{\beta} ,n_{\gamma} ,n_{\vartheta}) = (0, 1, 1, 0)$. The color gauge for the polarization $\vartheta$ and that for the orientation of the order parameter $\gamma$ are the same as those in the upper panels of Fig.~\ref{fig:4}. } 
	\label{fig:7}
\end{figure}

The circulation of polar-core half-quantum vortices can also be obtained from Eq.~(\ref{eq:circv_p}). 
Let us consider a polar-core vortex characterized by the winding number of $(n_{\alpha} ,n_{\beta} ,n_{\gamma} ,n_{\vartheta}) = (0, 1, 1, 0)$ shown in Fig.~\ref{fig:7}. 
The order parameter of this vortex is given by 
\begin{align} 
	\bm{\xi} = e^{i\varphi (R, \Phi )} \begin{pmatrix} 
	\sin {R} \cos { \frac{1}{2} \Phi} \\ 
	\sin {\frac{1}{2} \Phi} \\ 
	\cos {R} \cos { \frac{1}{2} \Phi} \end{pmatrix}, 
\end{align} 
which implies that the U($1$) gauge satisfies the condition $\varphi (R, \Phi + 2\pi ) - \varphi (R, \Phi ) = \pi$ and therefore the mass-current circulation is given by 
\begin{align} 
	\oint {\bm{v}} \cdot d\bm{l} = \frac{h}{2M}. 
\end{align} 

\section{\label{sec:6}Summary and Discussion} 
In this paper, we derive the su($3$) Mermin-Ho relation~(\ref{eq:rotv2}), which is expressed in terms of the Gell-Mann matrices and the su($3$) structure constants, and indicates nonholonomic structures of vortices in spin-$1$ BECs.  
For a BEC with fully-polarized spins, Eq.~(\ref{eq:rotv2}) reduces to the original Mermin-Ho relation in Eq.~(\ref{eq:MH}), which is expressed in terms of the so($3$) generators and structure constants. 

The local isomorphism between SO($3$) and SU($2$) implies that vortices dual to the CAT and MH vortices, both of which are described by Eq.~(\ref{eq:MH}), can be constructed from a set of three generators forming an su($2$) subalgebra of the su($3$) algebra. 
We show that three spin and nematicity observables, namely, $f_z$, $q_{xy}$, and $d_{x^2-y^2}$, belong to the other su($2$) subalgebra with the structure constant of $2$ and they form vortices dual to the spin vortices, i.e., the CAT and MH vortices, as visualized in Figs.~\ref{fig:3} and \ref{fig:4}.  
We also derive the formula to calculate the mass-current circulation around a vortex and identify the mass-current circulations of these spin-nematic vortices to be $3h / 2M$ (CAT) and $h/M$ (MH), respectively. 
The mass-current circulation for the dual CAT spin-nematic vortex in Fig.~\ref{fig:4} (i) is quantized in units of the half quantum $h/2M$, while the CAT vortex is characterized by the integer mass-current circulation of $2h/M$. 
On the other hand, the dual MH spin-nematic vortex has the same mass-current circulation as that of the MH vortex, both of which are given by $h/M$. 
The obtained formula to calculate the mass-current circulation around a vortex is applicable also to polar-core vortices. 

The vortices are not necessarily topologically stable in the enlarged order-parameter manifold~\cite{SKobayashi}; however, vortices such as the spin-nematic vortices can be realized by imposing appropriate boundary conditions. 
Such boundary conditions can be implemented by using a linearly polarized microwave, which can change the magnitude and the sign of the quadratic Zeeman energy~\cite{Gerbier,Leslie}.

\appendix 
\section*{ACKNOWLEDGEMENT} 
The authors would thank Michikazu Kobayashi, Yusuke Horinouchi, and Sho Higashikawa for useful discussions. 

\section{\label{as1}SU($3$) generators and structure constants} 
The generators of the SU(3) group are given by the Gell-Mann matrices: 
\begin{align} 
	& {\Lambda}_1 = \begin{pmatrix} 0 & 1 & 0 \\ 1 & 0 & 0 \\ 0 & 0 & 0 \end{pmatrix}, \ {\Lambda}_2 = \begin{pmatrix} 0 & -i & 0 \\ i & 0 & 0 \\ 0 & 0 & 0 \end{pmatrix}, \ {\Lambda}_3 = \begin{pmatrix} 1 & 0 & 0 \\ 0 & -1 & 0 \\ 0 & 0 & 0 \end{pmatrix}, \nonumber \\ 
	& {\Lambda}_4 = \begin{pmatrix} 0 & 0 & 1 \\ 0 & 0 & 0 \\ 1 & 0 & 0 \end{pmatrix}, \  {\Lambda}_5 = \begin{pmatrix} 0 & 0 & -i \\ 0 & 0 & 0 \\ i & 0 & 0 \end{pmatrix}, \ {\Lambda}_6 = \begin{pmatrix} 0 & 0 & 0 \\  0 & 0 & 1 \\ 0 & 1 & 0 \end{pmatrix}, \nonumber \\ 
	& {\Lambda}_7 = \begin{pmatrix} 0 & 0 & 0 \\  0 & 0 & -i \\ 0 & i & 0 \end{pmatrix}, \  {\Lambda}_8 = \frac{1}{\sqrt{3}} \begin{pmatrix} 1 & 0 & 0 \\ 0 & 1 & 0 \\ 0 & 0 & -2 \end{pmatrix}. \label{eq:gm3}
\end{align} 
The commutation relation between any two of them defines the structure constant $f_{ijk}$ through the relation 
\begin{align} 
	[{\Lambda}_i , {\Lambda}_j] = i \sum_{k=1}^8 f_{ijk} {\Lambda}_k. 
\end{align} 
The structure constants constitute a completely antisymmetric tensor of rank three. 
The nonzero components are 
\begin{align} 
\begin{cases} 
	& f_{123} = 2, \\ 
	& f_{147} = - f_{156} = f_{246} = f_{257} = f_{345} = - f_{367} = 1, \\ 
	& f_{458} = f_{678} = \sqrt{3},    
\end{cases} \label{eq:sf}
\end{align} 
and the components with permuted indices. 

\section{\label{as2}Derivation of Eq.~(\ref{eq:MH}) from Eq.~(\ref{eq:rotv2})} 
The Mermin-Ho relation can be obtained from Eq.~(\ref{eq:rotv2}), if we substitute the Gell-Mann matrices with spin and nematic components according to the correspondence in Table~\ref{tab:1}. 
It follows from the complete antisymmetry of the structure constant that Eq.~(\ref{eq:rotv2}) is equivalent to 
\begin{align} 
	\nabla \times \bm{v} = \frac{\hbar}{4M} \sum_{\mathrm{sgn} (ijk) = 1} f_{ijk} {\lambda}_i (\nabla {\lambda}_j ) \times (\nabla {\lambda}_k ), 
\end{align} 
where $\mathrm{sgn} (ijk)$ expresses the permutation parity of the subscripts $i$, $j$, and $k$. 
For each nonzero structure constant, the summation over the permutation of its suffixes can be simplified as follows:  
\begin{align} 
	& f_{123} {\lambda}_1 (\nabla {\lambda}_2 ) \times (\nabla {\lambda}_3 ) + f_{231} {\lambda}_2 (\nabla {\lambda}_3 ) \times (\nabla {\lambda}_1 ) \nonumber \\ 
	&+ f_{312} {\lambda}_3 (\nabla {\lambda}_1 ) \times (\nabla {\lambda}_2 ) \nonumber \\ 
	=& 2 [ q_{xy} (\nabla f_z ) \times (\nabla d_{x^2-y^2} ) + f_z (\nabla d_{x^2-y^2} ) \times (\nabla q_{xy} ) \nonumber \\ 
	&+ d_{x^2-y^2} (\nabla q_{xy} ) \times (\nabla f_z)] \nonumber \\ 
	=& (f_x^2 + f_y^2 ) [ f_x (\nabla f_y ) \times (\nabla f_z ) + f_y (\nabla f_z ) \times (\nabla f_x) \nonumber \\ 
	&+ 2 f_z (\nabla f_x) \times (\nabla f_y ) ],  \label{eq:b1}
\end{align} 
\begin{align} 
	& f_{147} {\lambda}_1 (\nabla {\lambda}_4 ) \times (\nabla {\lambda}_7 ) + f_{471} {\lambda}_4 (\nabla {\lambda}_7 ) \times (\nabla {\lambda}_1 ) \nonumber \\ 
	&+ f_{714} {\lambda}_7 (\nabla {\lambda}_1 ) \times (\nabla {\lambda}_4 ) \nonumber \\ 
	=& q_{xy} (\nabla q_{zx} ) \times (\nabla f_x) + q_{zx} (\nabla f_x) \times (\nabla q_{xy}) \nonumber \\ 
	&+ f_x (\nabla q_{xy}) \times (\nabla q_{zx}) \nonumber \\ 
	=& f_x^3 (\nabla f_y ) \times (\nabla f_z ),  
\end{align} 
\begin{align} 
	& f_{156} {\lambda}_1 (\nabla {\lambda}_5 ) \times (\nabla {\lambda}_6 ) + f_{561} {\lambda}_5 (\nabla {\lambda}_6 ) \times (\nabla {\lambda}_1 ) \nonumber \\ 
	&+ f_{615} {\lambda}_6 (\nabla {\lambda}_1 ) \times (\nabla {\lambda}_5 ) \nonumber \\ 
	=& - [ q_{xy} (\nabla f_y ) \times (\nabla q_{yz} ) + f_y (\nabla q_{yz}) \times (\nabla q_{xy}) \nonumber \\ 
	&+ q_{yz} (\nabla q_{xy}) \times (\nabla f_y) ] \nonumber \\ 
	=& f_y^3 (\nabla f_z ) \times (\nabla f_x ),  
\end{align} 
\begin{align} 
	& f_{246} {\lambda}_2 (\nabla {\lambda}_4 ) \times (\nabla {\lambda}_6 ) + f_{462} {\lambda}_4 (\nabla {\lambda}_6 ) \times (\nabla {\lambda}_2 ) \nonumber \\ 
	&+ f_{624} {\lambda}_6 (\nabla {\lambda}_2 ) \times (\nabla {\lambda}_4 ) \nonumber \\ 
	=& f_z (\nabla q_{zx} ) \times (\nabla q_{yz}) + q_{zx} (\nabla q_{yz}) \times (\nabla f_z) \nonumber \\ 
	&+ q_{yz} (\nabla f_z) \times (\nabla q_{zx}) \nonumber \\ 
	=& f_z^3 (\nabla f_x ) \times (\nabla f_y ),  
\end{align} 
\begin{align} 
	&f_{257} {\lambda}_2 (\nabla {\lambda}_5 ) \times (\nabla {\lambda}_7 ) + f_{572} {\lambda}_5 (\nabla {\lambda}_7 ) \times (\nabla {\lambda}_2 ) \nonumber \\ 
	&+ f_{725} {\lambda}_7 (\nabla {\lambda}_2 ) \times (\nabla {\lambda}_5 ) \nonumber \\
	=& f_x (\nabla f_y ) \times (\nabla f_z) + f_y (\nabla f_z ) \times (\nabla f_x) \nonumber \\ 
	&+ f_z (\nabla f_x ) \times (\nabla f_y),  
\end{align}
\begin{align} 
	&f_{345} {\lambda}_3 (\nabla {\lambda}_4 ) \times (\nabla {\lambda}_5 ) + f_{453} {\lambda}_4 (\nabla {\lambda}_3 ) \times (\nabla {\lambda}_5 ) \nonumber \\ 
	&+ f_{534} {\lambda}_5 (\nabla {\lambda}_3 ) \times (\nabla {\lambda}_4 ) \nonumber \\
	=&  - d_{x^2-y^2} (\nabla q_{zx} ) \times (\nabla f_y ) - q_{zx} (\nabla f_y) \times (\nabla d_{x^2-y^2} ) \nonumber \\ 
	&- f_y (\nabla d_{x^2-y^2} ) \times (\nabla q_{zx}) \nonumber \\ 
	=& \frac{1}{2} \{ f_x^2 [f_x (\nabla f_y ) \times (\nabla f_z ) + 2 f_y (\nabla f_z ) \times (\nabla f_x) \nonumber \\ 
	&+ f_z (\nabla f_x ) \times (\nabla f_y )] \nonumber \\ 
	&+ f_y^2 [ f_x (\nabla f_y ) \times (\nabla f_z ) - f_z ( \nabla f_x ) \times (\nabla f_y ) ] \},  
\end{align}  
\begin{align} 
	&f_{367} {\lambda}_3 (\nabla {\lambda}_6 ) \times (\nabla {\lambda}_7 ) + f_{673} {\lambda}_6 (\nabla {\lambda}_7 ) \times (\nabla {\lambda}_3 ) \nonumber \\ 
	&+ f_{736} {\lambda}_7 (\nabla {\lambda}_3 ) \times (\nabla {\lambda}_6 ) \nonumber \\
	=& - [ d_{x^2-y^2} (\nabla q_{yz} ) \times (\nabla f_x ) + q_{yz} (\nabla f_x) \times (\nabla d_{x^2-y^2} ) \nonumber \\ 
	&+ f_x (\nabla d_{x^2-y^2} ) \times (\nabla q_{yz}) ] \nonumber \\ 
	=& \frac{1}{2} \{ f_x^2 [f_y (\nabla f_z ) \times (\nabla f_x ) - f_z (\nabla f_x ) \times (\nabla f_y) ] \nonumber \\ 
	&+ f_y^2 [ 2 f_x (\nabla f_y ) \times (\nabla f_z ) + f_y ( \nabla f_z ) \times (\nabla f_x ) \nonumber \\ 
	&+ f_z (\nabla f_x ) \times (\nabla f_y ) ] \},   
\end{align} 
\begin{align}
	&f_{458} {\lambda}_4 (\nabla {\lambda}_5 ) \times (\nabla {\lambda}_8 ) + f_{584} {\lambda}_5 (\nabla {\lambda}_8 ) \times (\nabla {\lambda}_4 ) \nonumber \\ 
	&+ f_{845} {\lambda}_8 (\nabla {\lambda}_4 ) \times (\nabla {\lambda}_5 ) \nonumber \\ 
	=& \sqrt{3} [ q_{zx} (\nabla f_y ) \times (\nabla y) + f_y (\nabla y) \times (\nabla q_{zx} ) \nonumber \\ 
	&+ d_{3z^2-f^2} (\nabla q_{zx}) \times (\nabla f_y ) ] \nonumber \\ 
	=& \frac{1}{2} \{ f_x^2 [ f_x (\nabla f_y ) \times (\nabla f_z) + 2 f_y (\nabla f_z ) \times (\nabla f_x ) \nonumber \\ 
	&+ f_z (\nabla f_x ) \times (\nabla f_y ) ] \nonumber \\ 
	&+ f_y^2 [ - f_x (\nabla f_y ) \times (\nabla f_z ) + f_z ( \nabla f_x ) \times (\nabla f_y ) ] \} \nonumber \\ 
	&+ f_z^2 [f_x (\nabla f_y ) \times (\nabla f_z ) + 2 f_y (\nabla f_z ) \times (\nabla f_x ) \nonumber \\ 
	&+ f_z (\nabla f_x ) \times (\nabla f_y ) ], 
\end{align} 
and 
\begin{align}
	&f_{678} {\lambda}_6 (\nabla {\lambda}_7 ) \times (\nabla {\lambda}_8 ) + f_{786} {\lambda}_7 (\nabla {\lambda}_8 ) \times (\nabla {\lambda}_6 ) \nonumber \\ 
	&+ f_{867} {\lambda}_8 (\nabla {\lambda}_6 ) \times (\nabla {\lambda}_7 ) \nonumber \\ 
	=& \sqrt{3} [ - q_{yz} (\nabla f_x ) \times (\nabla y) - f_x (\nabla y) \times (\nabla q_{yz} ) \nonumber \\ 
	&- d_{3z^2-f^2} (\nabla q_{yz}) \times (\nabla f_x ) ] \nonumber \\ 
	=& \frac{1}{2} \{ f_x^2 [ - f_y (\nabla f_z ) \times (\nabla f_x) + f_z (\nabla f_x ) \times (\nabla f_y ) ] \nonumber \\  
	&+ f_y^2 [ 2 f_x (\nabla f_y ) \times (\nabla f_z ) + f_y ( \nabla f_z ) \times (\nabla f_x ) \nonumber \\ 
	&+ f_z ( \nabla f_x ) \times (\nabla f_y ) ] \} \nonumber \\ 
	&+ f_z^2 [2 f_x (\nabla f_y ) \times (\nabla f_z ) + f_y (\nabla f_z ) \times (\nabla f_x ) \nonumber \\ 
	&+ f_z (\nabla f_x ) \times (\nabla f_y ) ]. \label{eq:b2}
\end{align} 
Summing Eqs. (\ref{eq:b1})-(\ref{eq:b2}), we obtain 
\begin{align} 
	&\sum_{\mathrm{sgn} (ijk) = 1} f_{ijk} {\lambda}_i (\nabla {\lambda}_j ) \times (\nabla {\lambda}_k ) \nonumber \\ 
	=& 4 [ f_x (\nabla f_y ) \times (\nabla f_z ) + f_y (\nabla f_z ) \times (\nabla f_x ) \nonumber \\ 
	&+ f_z (\nabla f_x ) \times (\nabla f_y)  ]  \nonumber \\ 
	=& \frac{1}{2} \sum_{\mu \nu \lambda} f_{\mu} (\nabla f_{\nu} ) \times (\nabla f_{\lambda} ). 
\end{align} 
For the fully polarized case, i.e., $|\bm{f} |^2 = 1$, Eq.~(\ref{eq:rotv2}) reduces to the Mermin-Ho relation in Eq.~(\ref{eq:MH}).

\end{document}